\documentclass[aps,prl,twocolumn,showpacs,superscriptaddress]{revtex4}

\begin{document}


\title{Comment on Reply of Benedetto et al.}


\author{Dmitry V. Khmelev}
\email{D.Khmelev@newton.cam.ac.uk}
\affiliation{}  
\affiliation{Heriot-Watt
  University, Edinburgh, U.K. and Moscow State University, Russia}

\author{William J. Teahan}
\email{wjt@informatics.bangor.ac.uk}
\affiliation{University of Wales, Bangor, Dean Street, Bangor, LL57 1UT, U.K.}

\date{\today}

\begin{abstract}
  We regret to point out several inaccurate and misleading statements
that Benedetto {\em et al.} make in their Reply\cite{Bene:2003r} to
our Comment\cite{khmelev:2003c} on their paper titled ``Language Trees
and Zipping''\cite{Bene:2002}.
\end{abstract}


\maketitle

We regret to point out several inaccurate and misleading statements
that Benedetto {\em et al.} make in their Reply\cite{Bene:2003r} to
our Comment\cite{khmelev:2003c} on their paper titled ``Language Trees
and Zipping''\cite{Bene:2002}.

First they confusingly state in paragraph 7 that Russian and Greek
alphabets are not phonetic, putting Russian and Greek in a row with
Chinese, the latter enjoing hieroglyphic writings. Second, they use
unfair and irrelevant experiments in order to convince the reader that
the gzip-based approach is better than the Markov chains based
approach. Third, the figures reported for Newsgroups corpus seems to
be obtained on a randomly selected small subset of the Newsgroups corpus,
which probably makes them completely meaningless in the discussed
topic. Fourth, their reference to RAR compressor classification
performance for refuting our Comment is irrelevant to our Comment and
their Letter\cite{Bene:2002}. And fifth, authors of~\cite{Bene:2003r}
obviously experience some problems with scientific English language.
We elaborate on each of these points in more detail in the subsequent
paragraphs.

It is a well-established fact that Russian language as well as Greek
enjoys phonetic alphabet. Perhaps, Benedetto \emph{et
  al.}~\cite{Bene:2002} meant to use the transliteration for the
construction of Language Tree (LT). However this procedure has its
drawbacks like non-uniqueness, non-reversability, and inexactness of
the transformation. Most importantly this procedure requires some
\emph{knowledge} about the language, which shows that the requirement
for \emph{a-priori} information, pointed out in~\cite{khmelev:2003c}
remains valid contrary to the claim in~\cite{Bene:2002}.

We believe that if one wants to compare the perfomance of several
classification methods then the comparision should be performed in the
same experimental framework. To start with, let us denote by $M$, $G$,
$g$ the classification performance of the following methods,
respectively, on the corpus, discussed in~\cite{Khmelev:2000}: Markov
Chains approach~\cite{Khmelev:2000}; attribution with a single
source using gzip~\cite{Kukush:2001}; and attribution with
multiple-source using gzip~\cite{Bene:2002}. Let us denote by
$M'$, $G'$, $g'$ the classification performance of these methods in
the framework of~\cite{Bene:2002}, and, finally, let $M''$, $G''$,
$g''$ denote the classification performance of the same methods on the
Newsgroups dataset.  Notice that in~\cite{Bene:2002} only values for
$G''=60\%<g''=85\%$ and $G''=77\%<g''=93\%$ are presented. One can not
make any conclusion about $M'$ or $M''$ using these data, so our
statement about superiority of Markov Chains approach with respect to
gzip approach (either with a single- or multiple-source files)
remains valid.  Moreover, we have stated in our
Comment~\cite{khmelev:2003c} that $M=69/82\approx84\%$ is greater than
$G=50/82\approx61\%$. We also reported to editors of Phys.Rev.Lett.
in our answer to the referee report of Benedetto \emph{et al.} that
$g=53/82\approx65\%$, which can indeed be considered as an argument
for our claim that generally Markov chains are more attractive than
gzip-based approach.

In our opinion the ``slightly different method'' of \cite{Bene:2002}
should be considered as an approach to the design of the experiment,
which leads to an extremely slow classification speed especially in
the case of thousands of documents to classify, where a thousand
source documents makes prohibitive the really large experiment on
classification. This gives rise to the question of the validity of the
figures $G''=60\%$ and $g''=85\%$ outlined in~\cite{Bene:2003r}.
Traditionally, the precision of the classification method on the
Newsgroups is measured in the following way: one performs a random
10-fold or 5-fold split and reports the average results of
cross-validation.  Typical numbers reported are around
80\%~\cite{Teahan:2000}, with 82.1\% for PPM (Markov-based) approach.
It would be interesting to know the technique used
by\cite{Bene:2003r}, since even 5-fold split validation by their
method would require about
$5\times(18828/5)\times(4\times18828/5)\approx284\times10^6$ calls of gzip
compression program, which is prohibitive on conventional computers.
If one wants to apply a complete cross-validation as suggested
in~\cite{Bene:2002}, then one has to do even more
$18828^2\approx354\times10^6$ calls of gzip. We suspect that the
figures $G''=60\%$ and $g''=85\%$, outlined in~\cite{Bene:2003r}, are
obtained on a randomly selected small subset of the Newsgroups, are
subject to essential random variation, and hence $G''=60\%$ and
$g''=85\%$ should not be used for quantitative comparision of a
single- and multiple-source file setting.

Finally, in our Comment we stated that Markov chain approach as
reported in~\cite{Kukush:2001} is superior to LZ approach \emph{used
  in}~\cite{Bene:2002}. This statement was misinterpreted
in~\cite{Bene:2003r} as a general statement that LZ approach is
outperformed by the \emph{simple Markov chain} approach and Benedetto
\emph{et al.}\cite{Bene:2003r} easily refute the misinterpreted
statement using our own result on RAR~\cite{Kukush:2001}. The correct
generalization (and the only possible understanding in view of
references given) of our statement is: \emph{for any modification of
  LZ compression scheme there exists a modification of Markov Chain
  approach (PPM compression scheme), which outperforms LZ in
  classification} (this statement is similar to a well-known postulate
among specialists: any modification of LZ compression scheme can be
outperformed by a properly modified PPM compression scheme).  The
highly sophisticated, going far beyond the naive use of Ziv-Lempel
theory, algorithm of RAR, know-how of its creator Eugene Roshal,
should be compared with, for example, the the state-of-art PPMd
(PPMonstr) algorithm developed recently by Dmitry Shkarin. And we find
extremely interesting and scientifically valuable that the tough
first-order Markov chain produce results competitive to highly
sophisticated algorithms. As for the polemical comparision between
Markov Chain and RAR compressor by Benedetto \emph{et
  al.}\cite{Bene:2003r}, we find it irrelevant in the framework of
their paper~\cite{Bene:2002}. Indeed, if Benedetto \emph{et
  al.}\cite{Bene:2003r} stand for technical
details, like multiple- and single- source classification, they should
restrict their method to application of gzip only, which is the main 
technical detail of their Letter~\cite{Bene:2002}.

As a final remark we would like to point out that
Reply~\cite{Bene:2003r} exhibits some language
mistakes of it's authors themselves. Indeed, they reference to our
comment~\cite{khmelev:2003c} using expression ``Khmelev \emph{et al}''
as if~\cite{khmelev:2003c} has at least three co-authors (common
meaning of \emph{et al} is \emph{and others}).

To sum up, one can not draw any conclusion on the comparision between
$n$th order Markov chain approach (by which we
meant\cite{khmelev:2003c} the PPM approach as well) with gzip-based
approach from the statements, given in~\cite{Bene:2003r}. We also
believe that the authors of\cite{Bene:2003r} were not aware of our
reported figure for $g=64\%$; otherwise it looks very strange that
they did not mention this argument in their Reply. Also we suggest to
authors of~\cite{Bene:2003r} to present a fair comparision of their
method against others, \cite{Teahan:2000}, \cite{Kukush:2001} and,
e.g., SVM
approach~\cite{allwein00reducing,crammer00learnability,weston98multiclass}.

P.S. This story shows that editors of physical journal like
Phys.Rev.Lett. perhaps should avoid publishing papers
like~\cite{Bene:2002}, because Phys.Rev.Lett. referees do not have
enough experience to identify scientifical value and mistakes in
non-physical papers. We also encourage physicists and mathematicians
to send their non-physical and non-mathematical papers to appropriate
scientific journals, even if they are not so well-known as
Phys.Rev.Lett. Probably such a publication would not yield much
publicity, but the quality and scientific value of the paper would
increase significantly.

The example with~\cite{Bene:2002} is not unique. A similar story,
which reappears time-to-time in newspapers, is the story about
computing using DNA, described in details in~\cite{gusfield}. It is
possible to do computations with DNA. However, the amount of DNA,
required for solution of, say, salesmen problem with 100 cities, is
comparable with the Earth mass, which makes it's use impractical and
impossible. Notice that computer science methods allow to solve
practical salesmen problem in reasonable time for number of cities
like $10^6$ on contemporary computers.  However, the authors of DNA
computation speculative approach speculate that the effectiveness
issue will be solved in future, a strange analogy with suggestion
of~\cite{Bene:2003r}.

Notice also that a publication of non-physical paper in physical
journal evidence the crisis in physics, which responsible phisicists
should aware of. Otherwise why phisicists are publishing speculative
papers on non-physical subjects? Is not this the evidence that they
can not find application of their abilities in physics?

\bibliography{replycom}

\end{document}